\documentstyle[preprint,prl,aps,eqsecnum]{revtex}

\def\inbar{\vrule height1.5ex width.4pt depth0pt}
\def\IC{\relax\hbox{\kern.25em$\inbar\kern-.3em{\rm C}$}}
\def\IP{\relax{\rm I\kern-.18em P}}
\def\IF{\relax{\rm I\kern-.18em F}}
\def\IZ{\relax\ifmmode\hbox{Z\kern-.4em Z}\else{Z\kern-.4em Z}\fi}
\def\IR{\relax{\rm I\kern-.18em R}}

\def\I1{\relax{\rm 1\kern-.35em 1}}

\newcommand\NP{{\it Nucl. Phys.\ }}

\newcommand\PL{{\it Phys. Lett.\ }}
\newcommand\PR{{\it Phys. Rev.\ }}
\newcommand\PRL{{\it Phys. Rev. Lett.\ }}

\newcommand\Mod{{\it Mod. Phys. Lett.\ }}

\begin{document}
{\tighten 
\title{\bf Scattering of Four-Dimensional Black Holes}
\author{Jeremy Michelson}
\address{\vspace{3mm} Department of Physics \\ University of California \\
	Santa Barbara, California~~93106 \\ \vspace{3mm}
        {\rm and} \\ \vspace{3mm}
        Jefferson Physical Laboratories \\
        Harvard University \\
        Cambridge, Massachusetts~~02138 \\ \vspace{3mm}
        e-mail: jeremy@cosmic1.physics.ucsb.edu \\ \vspace{5mm}}
\date{August 25, 1997}
\preprint{UCSBTH-97-18, HUTP-97/A045, hep-th/9708091}
\maketitle
\begin{abstract}
The moduli space metric for an arbitrary number of extremal black holes
in four dimensions
with arbitrary relatively supersymmetric charges is found.
\end{abstract}

} 

\section{Introduction} \label{intro}
Black holes have proven to be an excellent testing ground for theories
of gravity.  In particular, one of the recent exciting developments
in string theory has been the reproduction of many of the macroscopic
black hole properties from the microscopic D-brane picture---for reviews
and references, see e.g.~\cite{horrev,maldrev}.  At the same time, one
of the current puzzles is the failed attempt in~\cite{dps} to
obtain, from a
microscopic calculation,
the macroscopic scattering of a D-string probe off a five-dimensional
supersymmetric black hole carrying the maximum three charges.  
Specifically, the interaction that is quadratic in
the charges was reproduced exactly, but the cubic term, which was seen
in~\cite{us} to be a degeneration of a three point interaction, was not
at all reproduced by the microscopic calculation.

In~\cite{us}, the macroscopic scattering of an arbitrary number of
the triply-charged supersymmetric five-dimensional black holes was given.
A proposal
for a microscopic calculation, based on the just-mentioned observation
of the origin of the three-point interaction, was also given.  In this paper,
scattering of supersymmetric four-dimensional black holes carrying four
charges
will be
discussed.  The motivation here rests on the fact that these
non-singular
black holes can be made purely out of D-brane~\cite{kleb,vjlars,vj,fermald}.
(If a supersymmetric four-dimensional black hole has fewer than four charges,
then it will be singular at the horizon.)
In principle, this makes the microscopic structure more
transparent~\cite{kleb,vjlars}.
This is in distinction to the five-dimensional case,
where despite requiring only three charges for non-singularity at the horizon,
there is no U-dual basis in which the charges are pure D-brane; the
usual description is as collections of parallel 5-branes and strings, with
momentum 
along the strings.  The difference in four dimensions is due to the additional
internal direction allowing the conditions for preservation of a
supersymmetry to be satisfied by a more general brane configuration.

In section~\ref{construct} we construct and discuss the black hole solution.
The black hole solution that we use is actually familiar from the heterotic
string---see e.g.~\cite{ct,ct2}.  We rederive the
solution in a way that makes explicit its Type II origin; this complements
the discussion in~\cite{cvj}.
In section~\ref{scatter} we give the effective action
that describes the scattering of several of these black holes, and give
a lengthy discussion of its U-dual generalization.  In particular, the
U-dual formulation of the three-point function is rather technical.
While we only explicitly calculate the effective action for 
black holes with four charges, we explain at the end of section~\ref{scatter} why
the U-duality invariant formula should hold for arbitrary supersymmetric
black holes including the black holes of~\cite{ct2} that carry five charges.
In section~\ref{conc} we conclude with a discussion of the scattering of
two black
holes.

\section{The Black Hole Solution} \label{construct}
In~\cite{vj}, black holes were constructed purely out of e.g.\ %
several D-4-branes,
intersecting at arbitrary $U(3)$ angles in the compact torus, and a D-0-brane.
In~\cite{kleb,vjlars,fermald},
the black holes were constructed out of e.g.\ orthogonally
intersecting D-3-branes.  However, because we will be doing the macroscopic
calculation, it will be convenient to use neither of these
descriptions in this paper.  Instead, we would like to find an NS-NS
description of the black holes, so we can use the formulas of~\cite{maharana,%
cvetic} for the dimensionally reduced supergravity lagrangian.
This can be obtained, for example, via the following series of dualities from
the D-3-brane configuration of~\cite{kleb,vjlars,fermald}:%
\footnote{It is also possible to obtain the NS-NS black hole from
the IIA NS-5-brane, D-6-brane and D-2-brane with momentum configuration
of~\cite{jmphd}.}
\begin{eqnarray} \label{dualities}
\mbox{\scriptsize
\begin{tabular}{ccc}
\begin{tabular}{c|llllllllll}
{IIB} & 0 & 1 & 2 & 3 & 4 & 5 & 6 & 7 & 8 & 9 \\ \hline
{D-3} & X &   &   &   & X & X & X &   &   &   \\
{D-3} & X &   &   &   &   &   & X & X & X &   \\
{D-3} & X &   &   &   & X &   &   &   & X & X \\
{D-3} & X &   &   &   &   & X &   & X &   & X
\end{tabular}
& {\Large $\stackrel{\mbox{\scriptsize T4,T5}}{\longrightarrow}$} &
\begin{tabular}{c|llllllllll}
{IIB} & 0 & 1 & 2 & 3 & 4 & 5 & 6 & 7 & 8 & 9 \\ \hline
{D-1} & X &   &   &   &   &   & X &   &   &   \\
{D-5} & X &   &   &   & X & X & X & X & X &   \\
{D-3} & X &   &   &   &   & X &   &   & X & X \\
{D-3} & X &   &   &   & X &   &   & X &   & X
\end{tabular}
\\
\end{tabular}} \nonumber \\ \mbox{ \scriptsize \begin{tabular}{ccc}
& & {\Large $\downarrow$} {S} \\
\begin{tabular}{c|llllllllll}
{IIB} & 0 & 1 & 2 & 3 & 4 & 5 & 6 & 7 & 8 & 9 \\ \hline
{NS-1}& X &   &   &   &   &   & X &   &   &   \\
{ETN} &   & X & X & X &   &   &   &   &   & X \\
{D-1} & X &   &   &   &   &   &   &   & X &   \\
{D-3} & X &   &   &   & X & X &   & X &   &  
\end{tabular}
& {\Large $\stackrel{\mbox{\scriptsize T5,T9}}{\longleftarrow}$} &
\begin{tabular}{c|llllllllll}
{IIB}  & 0 & 1 & 2 & 3 & 4 & 5 & 6 & 7 & 8 & 9 \\ \hline
{NS-1} & X &   &   &   &   &   & X &   &   &   \\
{NS-5} & X &   &   &   & X & X & X & X & X &   \\
{D-3}  & X &   &   &   &   & X &   &   & X & X \\
{D-3}  & X &   &   &   & X &   &   & X &   & X
\end{tabular}
\\
\end{tabular} } \\ \mbox{\scriptsize \begin{tabular}{ccc}
{T8,T6} {\Large $\downarrow$} \\
\begin{tabular}{c|llllllllll}
{IIB}  & 0 & 1 & 2 & 3 & 4 & 5 & 6 & 7 & 8 & 9 \\ \hline
{mom}  &   &   &   &   &   &   & X &   &   &   \\
{ETN}  &   & X & X & X &   &   &   &   &   & X \\
{D-1}  & X &   &   &   &   &   & X &   &   &   \\
{D-5}  & X &   &   &   & X & X & X & X & X &  
\end{tabular}
& {\Large $\stackrel{\mbox{\scriptsize S}} \longrightarrow$} &
\begin{tabular}{c|llllllllll}
{IIB}   & 0 & 1 & 2 & 3 & 4 & 5 & 6 & 7 & 8 & 9 \\ \hline
{mom}   &   &   &   &   &   &   & X &   &   &   \\
{ETN}   &   & X & X & X &   &   &   &   &   & X \\
{NS-1}  & X &   &   &   &   &   & X &   &   &   \\
{NS-5}  & X &   &   &   & X & X & X & X & X &  
\end{tabular}
\end{tabular} \nonumber 
} 
\end{eqnarray}
Note that under the T-duality in the 6-direction (T6), the fundamental
string parallel to the 6-direction transformed into a unit of Kaluza-Klein
momentum.  This is just the well-known momentum--winding exchange.
Recalling that the Kaluza-Klein monopole is essentially the product
of time and Euclidean Taub-NUT (ETN)~\cite{kkm1,kkm2},  and that
the NS-5-brane is the magnetic dual of the NS-string (c.f.\ equation~%
(\ref{hsoln10}) below)
the magnetic-dual of this phenomenon is the NS-5-brane--ETN transformation
under the T9 perpendicular to the NS-5-brane~\cite{town} (compare also
with~\cite{cvj}).

Now applying the harmonic function rule for
orthogonally intersecting branes in ten-
dimensions~\cite{tsh,gaunt,argurio}
gives (relabeling $9 \rightarrow 4$ and $6 \rightarrow 9$)
\begin{mathletters} \label{soln10}
\begin{eqnarray}
\label{gsoln10}
ds_{\mbox{\scriptsize} str}^2 & = & \psi_1^{-1} [-dt^2 + dx_9^2
  + \frac{Q_R}{r} (dt -dx_9)^2] 
  + \psi_5 \psi_E^{-1} (dx_4 + Q_E (1-\cos \theta) d \phi)^2 
\nonumber \\
& & + \psi_5 \psi_E (dr^2 + r^2 d \theta^2 + r^2 \sin^2 \theta d \phi^2)
    + dx_5^2 + dx_6^2 + dx_7^2 + dx_8^2, \\
\label{dilsoln10}
\varphi &=& \frac{1}{2} \ln (\psi_5 \psi_1^{-1}), \\
\label{hsoln10}
H &=& -Q_5 \sin \theta d \theta \wedge d\phi \wedge dx_4
     + \psi_1^{-2} \frac{d \psi_1}{d r} dt \wedge dr \wedge dx_9, \\
\label{psi1soln10}
\psi_1 &=& 1 + \frac{Q_1}{r}, \\
\label{psi5soln10}
\psi_5 &=& 1 + \frac{Q_5}{r}, \\
\label{psiEsoln10}
\psi_E &=& 1 + \frac{Q_E}{r}.
\end{eqnarray}
\end{mathletters}
Here we have postulated an obvious generalization of the harmonic function
rule to configurations involving the ETN; in particular, the ETN does
not contribute an overall conformal factor, in analogy to the Kaluza-Klein
momentum.
For notational simplicity, only the one-centred black hole has been written;
the generalization to the multi-black hole is almost obvious---see, e.g.~%
\cite{tseyt} for details on multi-centred ETN.  We have also set the
string coupling constant $g=e^{\varphi_\infty}=1$, where the subscript
denotes evaluation at spatial infinity.
The $Q_{\alpha}$s are constants; see also equations~(\ref{normq1})--(%
\ref{normqE}).
It is readily verified that equation~%
(\ref{soln10}) satisfies the equations of motion of the (string-frame)
NS-NS IIB action
\begin{equation} \label{action10}
S = \frac{1}{16 \pi G_{10}}
  \int d^{10}x \sqrt{-g}e^{-2 \varphi} [R + 4 (\nabla \varphi)^2
   - \frac{1}{12} H^2],
\end{equation}
where $G_{10} = 8 \pi^6 g^2 \alpha'^4$ is the ten-dimensional Newton constant.
This action, of course, describes the universal sector of all the 
string theories,
and, in fact, the solution of equation~(\ref{soln10}) is not new,
having been discussed in the context of the heterotic string in
e.g.~\cite{ct}.

Dimensional reduction on a $T^6$ now proceeds in the usual way~%
\cite{maharana}.  Of course, the NS-5-brane and the ETN give rise
to magnetic charges in 4-dimensions; it is therefore convenient to
dualize the corresponding vectors, and write the theory in terms of the
magnetic vector potentials and field strengths for which the
Bianchi identity and equation of motion are interchanged, e.g.\ %
$d \tilde{A}^{(2)}_4 \equiv \tilde{F}^{(2)}_4 =
e^{-2 \varphi} G^{44} \star F^{(2)}_4$, using the notation of~%
\cite{maharana}, and tildes to denote magnetic quantities.  The
four-dimensional, Einstein frame
action, and solution for multiple black holes carrying
four charges, are then,
\begin{eqnarray} \label{action4}
S &= & \frac{1}{16 \pi G_4}
   \int d^4x \sqrt{-g} \left\{ R - 2 (\partial_\mu \phi)^2
   - \frac{1}{4} (\partial_\mu \ln G_{44})^2 - \frac{1}{4} (\partial_\mu
   \ln G_{99})^2 - \frac{1}{4} e^{2 \varphi} G^{-1}_{44} (\tilde{F}^{(1)4}_{
   \mu \nu})^2 \right. \nonumber \\
& & \left. - \frac{1}{4} e^{-2 \varphi} G_{99} (F^{(1)9}_{\mu \nu})^2
   - \frac{1}{4} e^{2 \varphi} G_{44} (\tilde{F}^{(2)}_{4 \mu \nu})^2
   - \frac{1}{4} e^{-2 \varphi} G^{-1}_{99} (F^{(2)}_{9 \mu \nu})^2 \right\} ,
\end{eqnarray}
\begin{mathletters} \label{soln4}
\begin{eqnarray} \label{gsoln4}
ds_{\mbox{\scriptsize E}}^2 &=& -(\psi_1 \psi_5 \psi_R \psi_E)^{-\frac{1}{2}}
  dt^2 + (\psi_1 \psi_5 \psi_R \psi_E)^{\frac{1}{2}} d \vec{x}^2, \\
\label{dil4}
\varphi & = & \ln (\psi_1^{-\frac{1}{4}} \psi_5^{\frac{1}{4}} 
    \psi_R^{-\frac{1}{4}} \psi_E^{\frac{1}{4}}), \\
\label{g444}
G_{44} & = & \psi_5 \psi_E^{-1}, \\
\label{g994}
G_{99} & = & \psi_1^{-1} \psi_R, \\
\label{A1soln4}
A^{(2)}_9 & = & \psi_{1}^{-1} dt, \\
\label{A5soln4}
\tilde{A}^{(2)}_4 & = & \psi_{5}^{-1} dt, \\
\label{ARsoln4}
A^{(1)9} & = & \psi_{R}^{-1} dt, \\
\label{AEsoln4}
\tilde{A}^{(1)4} & = & \psi_E^{-1} dt,
\end{eqnarray}
\end{mathletters}
where the $\psi_\alpha$s, $\alpha \in \{1,5,R,E\}$ are harmonic functions,
\begin{mathletters} \label{psiandQ}
\begin{eqnarray} \label{psi}
\psi_\alpha &= &1 + \sum_{a=1}^N \frac{Q_{\alpha a}}{r_a}, \\
\label{normq1}
Q_{1a} &=& \frac{4 G_4 R_9}{\alpha'} n_{1a},\\
\label{normq5}
Q_{5a} &=& \frac{\alpha'}{2 R_4} n_{5a}, \\
\label{normqR}
Q_{Ra} &=& \frac{4 G_4}{R_9} n_{Ra},\\
\label{normqE}
Q_{Ea} &=& \frac{R_4}{2} n_{Ea},
\end{eqnarray}
\end{mathletters}
where the $n_{\alpha a}$ are non-negative integers.
$N$ is the number of black holes, and $\vec{r}_a$ is their positions.
The radii of the internal circles are $R_4,\ldots,R_9$ and
the four-dimensional Newton constant is $G_4 = \frac{g^2 \alpha'^4}%
{R_4 R_5 R_6 R_7 R_8 R_9}$.%
\footnote{For details on deriving the quantization of the charges and
the value of the $D$-dimensional Newton constant, see e.g.~\cite{jmphd}.  In
particular, we obtained the quantum of $Q_{E}$ by T-dualizing the
quantum of $Q_{5}$.}

\section{The Effective Action and U-duality} \label{scatter}
The Manton-type scattering calculation~\cite{man} proceeds exactly as 
in~\cite{fe,shir,us}, so we leave out
all the details here.  The result to ${\cal O}(\vec{v}^2)$
is
\begin{eqnarray} \label{result}
S_{\mbox{\scriptsize eff}} &=& \int dt  \left\{ -\sum_a m_a 
   + \frac{1}{2} \sum_a m_a \vec{v}_a^2 
   + \frac{1}{2 l_p^2} \sum_{\alpha<\beta} \sum_{a,b} Q_{\alpha a} Q_{\beta b}
     \frac{|\vec{v}_a - \vec{v}_b|^2}{r_{ab}} \right. \nonumber \\
& & + \left. \frac{1}{4 l_p^2} 
          \sum_{\stackrel{\mbox{\scriptsize $\alpha<\beta$}}
                   {\mbox{\scriptsize $\gamma \neq \alpha, \beta$}}}  
     \sum_{a,b,c} Q_{\alpha a} Q_{\beta b} Q_{\gamma c} |\vec{v}_a -
     \vec{v}_b|^2 (\frac{1}{r_{ab} r_{ac}} + \frac{1}{r_{ab}r_{bc}} -
     \frac{1}{r_{ac}r_{bc}}) \right. \nonumber \\
& & \left. + \frac{1}{2 l_p^2}  
    \sum_{\stackrel{\mbox{\scriptsize $\alpha<\beta; \gamma<\delta$}}
             {\mbox{\scriptsize $\alpha, \beta, \gamma, \delta$ all distinct}}}
     \sum_{a,b,c,d} 
Q_{\alpha a} Q_{\beta b} Q_{\gamma c} Q_{\delta d}
 |\vec{v}_a - \vec{v}_b|^2 
 \int d^3x \frac{\vec{r}_a \cdot \vec{r}_b}{4 \pi r_a^3 r_b^3 r_c r_d} 
 \right\},
\end{eqnarray}
where saturation of the Bogomol'nyi bound gives~\cite{jmphd}
\begin{equation} \label{bps}
m_a = \frac{1}{l_p^2} (Q_{1a} + Q_{5a} + Q_{Ra} + Q_{Ea}),
\end{equation}
and the four-dimensional Planck constant is $l_p = \sqrt{4 G_4} = \frac{%
g \alpha'^2}{\sqrt{2 R_4 \ldots R_9}}$.

Note that the
$a=b$ terms in the multiple sums clearly don't contribute.  Furthermore,
in the triple sum, the singular terms for $a=c$ or $b=c$ cancel, and in
the quadruple sum, the integral converges, even when two or more of
the coordinates coincide.
When one of the charges, say $Q_{Ea}$ vanishes for every black hole,
then equation~(\ref{result}) reduces to the result of~\cite{us} when the
latter is reduced from five to four dimensions, as required by the
arguments of~\cite{myers2}.

We would now like to make equation~(\ref{result}) U-duality invariant.
The U-dual expression for the terms linear and quadratic in the charges
follow exactly as in~\cite{dps,us}.  In particular the mass is
(in a sense elaborated below) already invariant, and the quadratic term
involves the masses and
contraction of two factors of the $E_{7(7)}$ charge vector
$q_{\Lambda a}$ with the inverse of the matrix of moduli, $({\cal M}_\infty^{
-1})^{\Lambda \Sigma}$ (see equation~(\ref{uresult}));
${\cal M}_{\Lambda \Sigma}$ is the matrix which
multiplies the kinetic term for the vector fields in the $E_{7(7)}$ 
invariant action.
The quartic term is clearly
proportional to the quartic invariant of the U-duality group~$E_{7(7)}$.
However, while in~\cite{us} the cubic term was proportional to the
cubic invariant of the five dimensional U-duality group~$E_{6(6)}$,
we can not directly associate such an interpretation to it in this case,
since $E_{7(7)}$ has no cubic invariant, nor does $E_{6(6)}$ imbed itself
into $E_{7(7)}$ in an intrinsically natural way.
Furthermore, it can be checked that we cannot use the invariants made out
of the matrix of moduli, the charge vectors and the masses to obtain the
cubic term; we can understand this because an expression involving
the matrix of moduli
could not reduce to the moduli independent $E_{6(6)}$ formula.%
\footnote{\label{overallM}
There will actually be overall matrix of moduli factors that
arise during the compactification from five to four dimensions.}

Instead, we recall,
following~\cite{kk,cvetic,feretal}, that there is an intrinsically natural way
of dissecting the $D=4, N=8$ central charge matrix.  Specifically, we
note that for each black hole
the moduli-dependent central charge matrix $\bbox{\sf Z}_a$ can be
$SU(8) \subset E_{7(7)}$
rotated into the form
\begin{equation} \label{ccharge}
\bbox{\sf Z}_a =
\mbox{diag} \{z_{1a},z_{2a},z_{3a},z_{4a}\} \otimes 
 \left( \begin{array}{cc} 0 & 1 \\ -1 & 0 \end{array} \right),
\end{equation}
with the $z_{\cdot a}$s the (possibly complex%
\footnote{But note that only the overall phase is invariant and not the
individual phases.}%
) ``eigenvalues''.
The largest eigenvalue, which we choose to be $z_{1a}$, is the mass of the 
$a$th
black hole, by the BPS condition.
In fact, since
$z_{1a} = l_p^2 m_a$, it, and more technically an $SU(2) \subset
SU(8)$, is singled out.  This was explained in~\cite{feretal} as the
$SU(2)$ corresponding to the supercharges (which transform linearly under
the $SU(8)$ automorphism) for which a complex linear combination annihilates
the state.  This is just the statement that it corresponds,
by the BPS condition, to the
unbroken supersymmetry.

In the case at hand,~\cite{cvetic,kk}
\begin{mathletters} \label{defzs}
\begin{eqnarray}
\label{defz1}
z_{1a} &=& Q_{1a} + Q_{5a} + Q_{Ra} + Q_{Ea} = l_p^2 m_a, \\
\label{defz2}
z_{2a} &=& Q_{1a} - Q_{5a} + Q_{Ra} - Q_{Ea}, \\
\label{defz3}
z_{3a} &=& Q_{1a} + Q_{5a} - Q_{Ra} - Q_{Ea}, \\
\label{defz4}
z_{4a} &=& Q_{1a} - Q_{5a} - Q_{Ra} + Q_{Ea}.
\end{eqnarray}
\end{mathletters}
It is easily checked that
\begin{eqnarray} \label{cubicu}
\sum_{\alpha \neq \beta \neq \gamma} Q_{\alpha a} Q_{\beta b} Q_{\gamma c}
& = & \frac{1}{16} \left\{ z_{1a} \left(z_{1b} z_{1c} - \sum_{I=2}^4
 z_{Ib} z^*_{Ic}\right) + (\mbox{5 perms}) \right\} \nonumber \\
& &  + \frac{1}{8} \left\{ \mbox{Re} (z_{2a} 
   z_{3b} z_{4b}) + (\mbox{5 perms}) \right\},
\end{eqnarray}
where again we have taken into account the fact that for the more general 
black holes, the $z_{\cdot a}$ are complex.  
Furthermore, if we set $\bbox{\sf Z}_a$ real and traceless, then we can
make contact with the real, traceless
five dimensional central charge matrix.   Specifically,
in this case equation~(\ref{cubicu})
is equivalent (up to a proportionality constant)
to the 5-dimensional $E_{6(6)}$ symmetric invariant,
written in the form~\cite{cvetic}
$\sum_{I=1}^4 z_{Ia} z_{Ib} z_{Ic}$.
In other words, when we restrict to black holes with three charges (or
fewer), then we recover the five-dimensional U-duality invariant formula.

Of course, we still need to convert equation~(\ref{cubicu}) into
a ``U-duality'' invariant formula involving the charge vectors $q_{\Lambda a}$.
The formula won't be truly U-duality invariant because we are decomposing
$E_{7(7)} \supset SU(8) \supset SU(2) \times SU(6)$; it will
only be invariant under the subgroup.%
\footnote{
As the $SU(8)$ is the maximal compact subgroup of $E_{7(7)}$,
this is (almost) the maximal decomposition of $E_{7(7)}$ involving our $SU(2)$
factor.  There is also a possible $U(1)$ factor; however, we have fixed the
$U(1)$ by demanding that $z_{1a}=l_p^2 m_a$, i.e.\ by fixing that $z_{1a}$ be
real.
}
The central charge matrix transforms linearly in the $\bbox{28} \oplus
\bbox{\overline{28}}$
of $SU(8)$;
under the above decomposition~\cite{pat,feretal},
\begin{equation} \label{decomp56}
\bbox{28} \oplus \bbox{\overline{28}} \rightarrow 
   (\bbox{1},\bbox{15})
   \oplus (\bbox{1},\bbox{\overline{15}})
   \oplus (\bbox{2},\bbox{6})
   \oplus (\bbox{2},\bbox{\overline{6}})
   \oplus (\bbox{1},\bbox{1})
   \oplus (\bbox{1},\bbox{1})
.
\end{equation}
Clearly, $z_{1a} \in (\bbox{1},\bbox{1})
$ and the other $z_{\cdot a} \in
 (\bbox{1},\bbox{15})
$.  So, the first term of equation~(\ref{cubicu})
is
\begin{mathletters} \label{justifyu}
\begin{equation} \label{firstjustu}
(\bbox{1},\bbox{1})
\otimes \{ [(\bbox{1},\bbox{1})]^2 +
 (\bbox{1},\bbox{15})
\otimes (\bbox{1},\bbox{\overline{15}})
\},
\end{equation}
which indeed contains a singlet, as required.
The second term of equation~(\ref{cubicu}) is
\begin{equation} \label{secondjustu}
[(\bbox{1},\bbox{15})]^3 + [(\bbox{1},\bbox{\overline{15}})]^3,
\end{equation}
\end{mathletters}
and again
each term contains a singlet.%
\footnote{This follows since the $\bbox{15} \in SU(6)$ is an antisymmetric
product of two fundamentals.  The antisymmetric product of six fundamentals
is clearly a singlet; this is the symmetric product of three~$\bbox{15}$s.
}
Note that this explains our choices of complex conjugation on
the right-hand side of equation~%
(\ref{cubicu}); any other polynomial choice that reduces to the left-%
hand side, and treats the $z_{\cdot a}$s and $z^*_{\cdot a}$s symmetrically%
\footnote{This is required since there is no invariant distinction between
the complex representations of $SU(6)$ and their complex conjugates.}%
,
would not be a singlet.  Thus, we have arrived at
equation~(\ref{cubicu}) uniquely.

So, to write down a more invariant expression we decompose the
integer-valued $E_{7(7)}$ charge vector $q_{\Lambda a}$.
More precisely, since we were working with the
central charge matrix, which is moduli dependent, it is convenient
to raise the index using the matrix of moduli:
\begin{equation} \label{raiseq}
q^{\Lambda}_a \equiv ({\cal M}_\infty^{-\frac{1}{2}})^{\Lambda \Sigma}
   q_{\Sigma a}.
\end{equation}
Then we can decompose $q^{\Lambda}_a$ as
$\{m_a, q^{A}_a, q^{\bar{A}}_a,\ldots\}$ where $m_a = l_p^{-2} z_{1 a}$
has been used for the $(\bbox{1},\bbox{1})$;
the index $A,\bar{A}=1,\ldots,15$ labels respectively
the $\bbox{15},
\bbox{\overline{15}} \in SU(6)$;
and the ellipses denote the representations that have not been included.
Then, we finally have the U-duality invariant version of
equation~(\ref{result}).
\begin{eqnarray} \label{uresult}
S_{\mbox{\scriptsize eff}} & = & 
  \int dt \left\{ - \sum_a m_a
  +\frac{1}{2} \sum_a m_a \vec{v}_a^2
  +\frac{1}{2}
       \sum_{a<b} (l_p^2 m_a m_b - q_{\Lambda a} ({\cal M}_\infty^{-1})^{
       \Lambda \Sigma}
       q_{\Sigma b}) \frac{|\vec{v}_a - \vec{v}_b|^2}{
       r_{ab}} \right. \nonumber \\
& & \left.
  +\frac{3}{32} \sum_{a<b} \sum_c \left[l_p^4 m_a m_b m_c - \frac{l_p^2}{6}
       (m_a q^{A}_{b} \delta_{A \bar{A}} q^{\bar{A}}_{c} 
           + \mbox{5 perms})
      + l_p d_{(6)ABC} 
           q^{A}_{a} q^{B}_{b} q^{C}_{c} \right. \right. \nonumber \\
& & \left. \left. \hspace{1.5cm}
      + l_p d^*_{(6)\bar{A}\bar{B}\bar{C}} 
      q^{\bar{A}}_a q^{\bar{B}}_b q^{\bar{C}}_c \right]
      |\vec{v}_a-\vec{v}_b|^2 
      \left[ \frac{1}{r_{ab} r_{ac}} +
      \frac{1}{r_{ab} r_{bc}} - \frac{1}{r_{ac}r_{bc}} \right] \right.
      \nonumber \\
& & \left.
  + \frac{l_p^2}{4} \sum_{a<b} \sum_{c,d} d^{\Lambda \Sigma \Gamma \Pi}
      q_{\Lambda a} q_{\Sigma b} q_{\Gamma c} q_{\Pi d}
      |\vec{v}_a - \vec{v}_b|^2 \int d^3x \frac{\vec{r}_a \cdot
      \vec{r}_b}{4 \pi r_a^3 r_b^3 r_c r_d} \right\}.
\end{eqnarray}
Here, $d_{(6)ABC}$ is proportional to the symmetric cubic invariant for the 
$\bbox{15}
\in SU(6)$, and $d^{\Lambda \Sigma \Gamma \Pi}$ is proportional to
the $E_{7(7)}$
cubic invariant.

Two final comments are required regarding the decomposition $E_{7(7)}
\supset SU(8) \supset SU(2) \times SU(6)$.  
First, it appears that we have assumed that the central charge
matrices for the black holes can be simultaneously diagonalized (in
the sense of equation~(\ref{ccharge})).
However, all we really
need to assume is that they can be simultaneously block-diagonalized
into $SU(2)$ and $SU(6)$ subgroups; our final expression, equation~%
(\ref{uresult}) is $SU(6)$ invariant and so does not require that the
matrices be diagonal.  That the block-diagonalization is possible is simply
the statement that the black holes preserve a common supersymmetry.
Incidentally, the block-diagonalization implies that the charges that
transform in the $(\bbox{2},\bbox{6})$ representation of $SU(2)\times SU(6)$
vanish; this is why there is no $(\bbox{1},\bbox{1})\otimes(\bbox{2},
\bbox{6})\otimes (\bbox{2},\bbox{\overline{6}})$ term in equation~%
(\ref{cubicu}) or~(\ref{uresult}).

Second, it
was implicitly assumed in the discussion
that the solution preserves exactly
$\frac{1}{8}$ of the supersymmetry.  If the solution preserves more
supersymmetry---i.e. if more than one $z_{\cdot a} = l_p^2 m_a$---then
there is no longer a natural $SU(2) \subset SU(8)$ but rather a larger
subgroup that is selected.  Nevertheless, it is easy from equation~%
(\ref{cubicu}) to see that no matter
how one chooses the $SU(2) \subset G$ (where, for a solution
preserving $\frac{1}{4}$ of the supersymmetry, $G=SU(4)$, for example)
one obtains the same answer for the cubic, namely zero, so there is no
ambiguity when there is more supersymmetry.

We now claim that equation~(\ref{uresult}), which was really only
derived for the special case of black holes with four charges, holds
for general (e.g.\ five charge) supersymmetric black hole configurations.
The forms of the two-point, three-point and four-point functions
have already been fixed uniquely by equation~(\ref{result}) and
the duality symmetries%
, so the only possible modification with the
additional charges, is the appearance of higher-point functions.  Since
these higher-point functions vanish when only four charges are
non-zero, they must be proportional to all five separate charges, and hence
cannot be made out of invariants of order less than five.  But the group
theory that led us to the ``U-duality'' invariant form for the cubic term
shows us that there are no such candidate invariants: all we can
work with 
is the $(\bbox{1},\bbox{15})$
and its complex conjugate, and since
cubing one gives a singlet, and multiplying one by the other gives a singlet
we can never get a higher-order invariant.  This leaves
the $E_{7(7)}$ invariants, and the only symmetric one is the quartic.
Thus, equation~(\ref{uresult}) is the general result.

\section{Discussion} \label{conc}
We have given the effective action to ${\cal O}(\vec{v}^2)$
for scattering of an arbitrary number of charged supersymmetric
four dimensional black holes.  The U-duality invariant form required a
technical discussion of the
natural $SU(2)\times SU(6)$ decomposition of the four-dimensional
U-duality group $E_{7(7)}$.  We would now like to give a slightly
more detailed discussion of the asymptotically flat
moduli space for two black holes.  From equation~%
(\ref{uresult}), the moduli space is
\begin{mathletters} \label{mod2bh}
\begin{eqnarray}
\label{mod2bhmetric}
ds^2 &=& \frac{1}{2} f(\vec{r}) (dr^2 + r^2 d\Omega^2),
\end{eqnarray}
where
\begin{eqnarray}
\label{mod2bhf}
f(\vec{r}) &=& \mu + \frac{\Gamma_{II}}{r} + \frac{\Gamma_{III}}{r^2} +
   \frac{\Gamma_{IV}}{r^3}
, \\
\label{mod2bhg2}
\Gamma_{II} &=& l_p^2 M \mu - q_{\Lambda 1} ({\cal M}_\infty^{-1})^%
        {\Lambda \Sigma} q_{\Sigma 2}, \\
\label{mod2bhg3}
\Gamma_{III} &=& \frac{3}{16} \left\{l_p^4 M^2 \mu 
     + l_p d_{(6)ABC} q^A_1 q^B_2 (q^C_1 + q^C_2)
     + l_p d^*_{(6)\bar{A}\bar{B}\bar{C}} q^{\bar{A}}_1 q^{\bar{B}}_2 
           (q^{\bar{C}}_1 + q^{\bar{C}}_2)
\right. \nonumber \\ & & \left.
     - \frac{l_p^2}{3} \left( M q^A_1 \delta_{A\bar{A}} q^{\bar{A}}_2
          + M q^A_2 \delta_{A\bar{A}} q^{\bar{A}}_1
          + m_1 q^A_2 \delta_{A\bar{A}} q^{\bar{A}}_2
          + m_2 q^A_1 \delta_{A\bar{A}} q^{\bar{A}}_1 \right) \right\}, \\
\label{mod2bhg4}
\Gamma_{IV} &=& \frac{l_p^2}{6} d^{\Lambda \Sigma \Gamma \Pi}
     q_{\Lambda 1} q_{\Sigma 2} \left(q_{\Gamma 1} q_{\Pi 1} + q_{\Gamma 2}
     q_{\Pi 2} \right),
\end{eqnarray}
and the centre of mass (relative mass) is $M=m_1+m_2$ 
($\mu=\frac{m_1 m_2}{m_1+m_2}$); also the relative coordinate is $\vec{r}=
\vec{r_2}-\vec{r_1}$.  In equation~(\ref{mod2bhmetric}), we have subtracted
away the centre of mass motion.  We have also omitted the term
\begin{equation} \label{omitmod2bhf} 
\frac{\pi^3 l_p^4}{4} d^{\Lambda \Sigma \Gamma \Pi}
     q_{\Lambda 1} q_{\Sigma 1} q_{\Gamma 2} q_{\Pi 2} \delta^{(3)}(\vec{r}),,
\end{equation}
\end{mathletters}
from equation~(\ref{mod2bhf}) because this contact interaction
only occurs at $r=0$ by which point
the moduli space approximation has presumably broken down. 
We recall that the moduli space approximation is only valid for small
velocities
and neglects radiation; in particular, in~\cite{fe} it was argued that the 
moduli
space approximation breaks down for $r \lesssim v_\infty^2 M$.

The coordinate singularity at $r=0$ of equation~%
(\ref{mod2bhmetric}) is removed, when $\Gamma_{IV} \neq 0$,
by performing the coordinate
transformation $\xi=-2 \frac{\alpha'^{\frac{3}{4}}}{\sqrt{r}}$.
One then finds that as $r \rightarrow 0$,  ($\xi \rightarrow \infty$),
there is a second asymptotic region that is conical with deficit angle~%
$\pi$~\cite{fe}.  For $\Gamma_{IV} = 0$ but $\Gamma_{III} \neq 0$,
the coordinate singularity at
$r=0$ is removed by the coordinate transformation $\xi = \ln \frac{r}{\sqrt{%
\alpha'}}$ 
to find that the
asymptotic region has topology $\IR \times S^2$.
If both $\Gamma_{IV} = 0$  and $\Gamma_{III} = 0$, then one removes the
$r=0$ coordinate singularity via $\xi = \sqrt{\sqrt{\alpha'} r}$ to again find
a second asymptotic region that is conical with deficit 
angle~$\pi$~\cite{shir}.
Thus, geodesics which extend to $r=0$ enter this second
asymptotic region; i.e.\ the black holes coalesce~\cite{fe,shir,dps}.
(It might seem strange that, having just rejected the contact interaction
for being at $r=0$, we are exploring the $r \rightarrow 0$ behaviour.
The point is that we can use equations~(\ref{mod2bh}), even as
$r \rightarrow 0$ (but $r \neq 0$), by taking
$v_\infty \rightarrow 0$.)

By examining
the geodesic equation as in~\cite{dps}, we find that the
turning point $r_c$ in the black hole motion is real and positive when
the impact parameter $b$ obeys
\begin{eqnarray} \label{bc}
b^2 &>& b_c^2 \equiv \mbox{\small $- \frac{\Gamma_{II}^2}{12 \mu^2}  + 
     \frac{\Gamma_{III}}{\mu} 
$}\nonumber \\ && \mbox{\small $
     + \frac{{\Gamma_{II}^4 + 18\,\Gamma_{II}\,\Gamma_{IV}\,{{\mu}^2}}}{
          {{12 \left( -\left( {\Gamma_{II}^6}\,{{\mu}^6}
                \right)  + 
             540\,{\Gamma_{II}^3}\,\Gamma_{IV}\,
              {{\mu}^8} + 
             24\,\left( 243\,{\Gamma_{IV}^2}\,
                 {{\mu}^{10}} + 
                {\sqrt{3}}\,
                 {\sqrt{\Gamma_{IV}\,{{\mu}^{14}}\,
                     {{\left( -{\Gamma_{II}^3} + 
                      27\,\Gamma_{IV}\,{{\mu}^2} \right) }^3
                      }}} \right)  \right) }^{{\frac{1}{3}}}}}
$}  \\ && \mbox{\small $
      + \frac{1}{12} 
        {{\left( -\frac{ {\Gamma_{II}^6}}{\mu^6}  + 540\,
           \frac{{\Gamma_{II}^3}\,
           \Gamma_{IV}}{{\mu}^4} + 
          24\,\left( 243\,\frac{\Gamma_{IV}^2}{\mu^2} + 
             {\sqrt{3}}\,{\sqrt{\frac{\Gamma_{IV}}{{\mu}^{10}}\,
                  {{\left( -{\Gamma_{II}^3} + 
                      27\,\Gamma_{IV}\,{{\mu}^2} \right) }^3
                    }}} \right)  \right) }^{{\frac{1}{3}}}}$}\nonumber.
\end{eqnarray}
In particular there is coalescence for $b \leq b_c$.  Na\"{\i}vely,
$b_c$ is only well-defined if either $\Gamma_{IV}=0$ or $\Gamma_{II}^3
\leq 27 \Gamma_{IV}  \mu^2$;
in fact, one must merely be careful
about how one chooses the cube roots in equation~(\ref{bc}).

As equation~(\ref{bc}) is rather obscure, we point out some special cases.
If we consider the black holes of section~\ref{construct}, with
$Q_{1a} = Q_{5a} = Q_{Ra} = Q_{Ea} = l_p^2 \frac{m_a}{4}$, then we have the
Reissner-Nordstr\"{o}m black holes of~\cite{fe}%
\footnote{Equation~(\ref{bc}) corrects the polynomial equation
for $b_c$ that was given in the reference.}
for which the right-hand side of equation~(\ref{bc}) is real and positive,
and so there can be coalescence.  Note that this is despite the
fact that for Reissner-Nordstr\"{o}m black holes,
$27 \Gamma_{IV} \mu^2 - \Gamma_{II}^3 < 0$.
For two Reissner-Nordstr\"{o}m black
holes of equal mass ($m_1 = m_2$), $b_c \approx 2.3660 \, l_p^2
\frac{M}{4}$, in agreement with~\cite{fe}.

If the black holes only carry three charges, then $\Gamma_{IV} = 0$;
we find $b_c = \sqrt{\frac{\Gamma_{III}}{\mu}}$.%
\footnote{To obtain this from equation~(\ref{bc}) requires
choosing $(-1)^\frac{1}{3} = \frac{1}{2} \pm i \frac{\sqrt{3}}{2}$.}
In fact, if the black holes carry fewer than three charges, so that also
$\Gamma_{III}=0$, then there is never coalescence (except for the obvious
case $b=0$); this is in agreement with the results of~\cite{shir}.
This is different from higher dimensions, where there is always a
critical, non-zero impact parameter below which there is coalescence~%
\cite{shir,us}.

\acknowledgments
I thank Vijay Balasubramanian, Eric Gimon, Gary Horowitz, John Pierre,
Joe Polchinski, Andy Strominger and Haisong Yang
for useful discussions.  I also thank David Kaplan for
collaboration on an earlier, related paper.
I am grateful to Harald H. Soleng for making~\cite{cartan}
available, which was very useful in the computation.
The hospitality of the Physics Department at Harvard University is appreciated.
Financial support from NSERC and NSF is gratefully acknowledged;
this work was also supported in part by DOE Grant No. DOE-91ER40618.


\begin{thebibliography}{99}

\bibitem{horrev} G.\ Horowitz, to appear in the proceedings of the Pacific
Conference on Gravitation and Cosmology, Seoul, Korea, February 1-6, 1996;
gr-qc/9604051.

\bibitem{maldrev} J.\ Maldacena, hep-th/9705078.

\bibitem{dps}{M.\ Douglas, J.\ Polchinski and A.\ Strominger,
hep-th/9703031.}

\bibitem{us}D.\ M.\ Kaplan and J.\ Michelson, hep-th/9707021.

\bibitem{kleb} I.\ R.\ Klebanov and A.\ A.\ Tseytlin, \NP {\bf B475}
(1996) 179; hep-th/9604166.

\bibitem{vjlars} V.\ Balasumbramanian, F.\ Larsen, \NP {\bf B478} 
(1996) 199; hep-th/9604189.

\bibitem{vj} V.\  Balasubramanian, F.\  Larsen and R.\ G.\ Leigh,
    hep-th/9704143.

\bibitem{fermald} S.\ Ferrara and J.\ Maldacena, hep-th/9706097.

\bibitem{ct} M.\ Cveti\v{c} and A.\ Tseytlin, \PL {\bf B366} (1996) 95;
    hep-th/9510097.

\bibitem{ct2} M.\ Cveti\v{c} and A.\ A.\ Tseytlin, \PR {\bf D53} (1996)
   5619; erratum \PR {\bf D55} (1997) 3907; hep-th/9512031.

\bibitem{cvj} C.\ V.\ Johnson, R.\ R.\ Khuri and R.\ C.\ Myers,
   \PL {\bf B378} (1996) 78; hep-th/9603061.

\bibitem{maharana}{J.\ Maharana and J.\ H.\ Schwarz, \NP
{\bf B390} (1993) 3; hep-th/9207016.}

\bibitem{cvetic} M.\ Cveti\v{c} and C.\ Hull, \NP {\bf B480} (1996) 296;
hep-th/9606193.

\bibitem{jmphd}{ J.\ Maldacena, Ph.D.\ Thesis, Princeton University;
    hep-th/9607235.}

\bibitem{kkm1} R.\ D.\ Sorkin, \PRL {\bf 51} 87.

\bibitem{kkm2} D.\ J.\ Gross and M.\ J.\ Perry, \NP {\bf B226} (1983) 29.

\bibitem{town} P.\ Townsend, lectures at the Cargese '97 ASI on Strings,
   Branes and Duality.

\bibitem{tsh} A.\ Tseytlin, \NP {\bf B475} (1996) 149; hep-th/9604035.

\bibitem{gaunt} J.\ P.\ Gauntlett, D.\ A.\ Kastor and J.\ Traschen,
\NP {\bf B478} (1996) 544; hep-th/9604179.

\bibitem{argurio} R.\ Argurio, F.\ Englert and L.\ Houart, \PL {\bf B398}
    (1997) 61; hep-th/9701042.

\bibitem{tseyt}{A.\ Tseytlin, \Mod {\bf A11} (1996) 689; hep-th/9601177.}

\bibitem{man}{N.\ S.\ Manton, \PL {\bf B110} (1982) 54.}

\bibitem{fe}{R.\ Ferrell and D.\ Eardley, \PRL {\bf 59} (1987) 1617.}

\bibitem{shir}{K.\ Shiraishi, \NP {\bf B402} (1993) 399.}

\bibitem{myers2} R.\ R.\ Khuri and R.\ C.\ Myers, \NP {\bf B466} (1996) 60;
   hep-th/9512061.

\bibitem{kk} R.\ Kallosh and B.\ Kol, \PR {\bf D53} (1996) 5344;
   hep-th/9602014.

\bibitem{feretal} L.\ Andrianopoli, R.\ D'Auria, S.\ Ferrara,
   P. Fr\'{e} and M. Trigiante, hep-th/9707087.

\bibitem{pat} J.\ Patera and D.\ Sankoff, {Tables of Branching
Rules for Representations of Simple Lie Algebras} (Les Presses de
l'Universit\'{e} de Montr\'{e}al, Montreal, 1973).

\bibitem{cartan}H.\ H.\ Soleng, {\it CARTAN: A Mathematica package for
tensor analysis}, gr-qc/9502035.

\end{thebibliography}
\end{document}